\documentclass[runningheads]{llncs}
\usepackage[T1]{fontenc}
\usepackage{graphicx}
\usepackage{tikz,pgfplots}
\usetikzlibrary{patterns}
\usepackage{cleveref}
\usepackage{cite}
\usepackage{multirow}

\begin{document}
\title{ISS-Scenario: Scenario-based Testing in CARLA}
%
%
\author{Renjue Li\inst{1}\orcidID{0000-0003-2472-0021} \and
Tianhang Qin\inst{1}\orcidID{0009-0003-5499-4581} \and
Cas Widdershoven\inst{1,2}\orcidID{0000-0002-1676-0747}}
\authorrunning{R. Li et al.}
%
\institute{Institute of Software, Chinese Academy of Sciences, Beijing, China \and Institute of Intelligent Software, Guangzhou, Guangzhou, P.\@ R.\@ China}
\maketitle              
\begin{abstract}
The rapidly evolving field of autonomous driving systems (ADSs) is full of promise. However, in order to fulfil these promises, ADSs need to be safe in all circumstances. This paper introduces ISS-Scenario, an autonomous driving testing framework in the paradigm of scenario-based testing. ISS-Scenario is designed for batch testing, exploration of test cases (e.g., potentially dangerous scenarios), and performance evaluation of autonomous vehicles (AVs). ISS-Scenario includes a diverse simulation scenario library with parametrized design. Furthermore, ISS-Scenario integrates two testing methods within the framework: random sampling and optimized search by means of a genetic algorithm. Finally, ISS-Scenario provides an accident replay feature, saving a log file for each test case which allows developers to replay and dissect scenarios where the ADS showed problematic behavior.

\end{abstract}
\section{Introduction}
Recent years have seen big improvements in the area of autonomous driving systems (ADSs). Advances in machine learning and computer vision have enabled higher and higher degrees of autonomy to ADSs, and businesses such as Tesla and BYD show that ADSs have reached a high enough state of maturity that they can be an important basis for commercial success. ADSs have the potential to transform society, but before that is the case there are still significant hurdles to be overcome.

One of the biggest hurdles is the question of safety. Modern traffic has to be safe in a variety of challenging situations, which is why drivers have to train and take an exam before being allowed on the road. Wide-spread adoption of ADSs would be contingent on them being at least as safe as human drivers, and ideally even safer. Hence, safety for ADSs is paramount.

In classical settings, safety of a system is shown either through model-based analysis, or through experimentation. However, for autonomous driving systems, models are inaccurate and too large to prove safety through model-based analysis. Moreover, real-world experimentation requires ADSs to drive millions if not billions of miles through many real-world scenarios, making this approach prohibitively expensive as well. For this reason, an important part in proving safety of ADSs is virtual, scenario-based testing. This type of scenario-based testing employs driving simulators such as CARLA to test an ADS in various different driving scenarios.

In this paper we introduce ISS-Scenarios. ISS-Scenarios is a testing framework for scenario-based testing of ADSs in the open-source CARLA simulator. It is designed for batch testing, exploration of test cases, and performance evaluation of autonomous vehicles. Unlike other testing frameworks, ISS-Scenarios includes a diverse simulation scenario library with parameterized design. Users can easily select suitable scenarios, configure parameter ranges (e.g., adjust weather or other traffic participants' positions and speeds), and automate testing to explore numerous test cases based on their needs.

Furthermore, ISS-Scenarios integrates two testing methods within the framework: random sampling and optimized search by means of a genetic algorithm. Due to its modular design, other sampling or search algorithms can be easily integrated into the framework. 

Finally, ISS-Scenarios provides an accident replay feature, saving a log file for each test case. Users can use the provided interface to replay the entire accident process in the simulator, enabling analysis of accidents using all simulator data.

The main contributions of this paper are as follows:
\begin{enumerate}
\item Development and design of a parameterized scenario library based on the CARLA simulator.
\item Design and implementation of a batch simulation scenario testing framework based on the parameterized scenario library. The goal is to create a simple and user-friendly simulation testing tool for testing, evaluating, and improving autonomous driving systems.
\item Using ISS-Scenarios, we select four scenarios from the scenario library, apply random sampling and optimized search within the testing framework, and test the response capabilities of the TCP and InterFuser autonomous driving systems in scenarios such as intersection encounters and pedestrian crossings. Experimental results indicate that despite the good performance of TCP and InterFuser in CARLA's benchmark tests, ISS-Scenarios can still effectively and automatically find a series of settings that ead to dangerous behavior.
\end{enumerate}

The paper has the following structure: in \Cref{sec:relworks} we will discuss related literature. In \Cref{sec:scenlib} we will give an overview of the different types of scenarios provided by ISS-Scenario. In \Cref{sec:batch} we will discuss the batch testing capabilities of ISS-Scenario, and in \Cref{sec:case} we will discuss some case studies done using ISS-Scenario. Finally we conclude in \Cref{sec:conc}.

ISS-Scenario is available at \url{https://github.com/CAS-LRJ/ISS\_Scenario}.

\section{Related Works}\label{sec:relworks}

Autonomous Driving Systems (ADS) \cite{interfuser, Transfuser, Pylot, LBC, LAV, TCP, ToromanoffWM20} demonstrate promising potential for efficient and safe driving pipelines. The modular ADS pipeline \cite{Pylot} comprises individual modules with distinct functionalities, such as perception, prediction, localization, planning, and control. Conversely, modern end-to-end pipelines \cite{interfuser, Transfuser, LBC, LAV, TCP} are presented as single comprehensive models that directly generate control signals from sensor inputs.
In our case study, we selected two state-of-the-art ADS, namely Interfuser \cite{interfuser} and TCP \cite{TCP}, to evaluate our testing framework. Real-world testing of ADS is prohibitively expensive, thus autonomous driving simulators \cite{BeamNG, CARLA} have been developed to facilitate large-scale testing before implementation on the road. We implemented our framework using the most popular simulator, CARLA \cite{CARLA}.
Previous works such as Scenic \cite{Scenic}, CommonRoad \cite{CommonRoad}, and GeoScenario \cite{GeoScenario} describe autonomous driving scenarios using domain-specific languages (DSLs) and utilize generated scenarios to test ADS. In our framework, we construct common corner-case scenarios using the CARLA ScenarioRunner, which streamlines the complex DSL definition process and enhances framework usability.
Search-based testing approaches \cite{DreossiFGKRVS19, 0008P00Y22, ZhongKR23, AbdessalemNBS18, calo2020generating, borg2021digital, gambi2019automatically, gambi2019asfault, tian2022mosat, haq2022efficient, AbdessalemPNBS18, KluckZWN19, LiLJTSHKI20, ArcainiZI21, Gladisch0HOVP19, Ishikawa20, LuoZAJZIW022} play a crucial role in testing ADS and identifying dangerous corner cases. We have adopted commonly used genetic algorithms to search for corner cases within our framework.





\section{ISS-Scenario Scenario Library}\label{sec:scenlib}
Central to ISS-Scenario is the scenario library. This extensive library contains many different scenarios an ADS may encounter, all of which can be adjusted through different parameters such as the speed of vehicles, or the positions of different actors on the road. This highly configurable approach allows for batch testing of ADSs in many different settings. There are five main categories of scenarios in ISS-Scenario:
\begin{itemize}
    \item \textbf{Obstacle recognition:} these are scenarios in which the ADS encounters static obstacles in the road. Examples of these include general static obstacles on a straight road, stationary vehicles, and obstacles after a turn at an intersection.
    \item \textbf{Intersection encounters:} these are scenarios in which the ADS has to deal with traffic at an intersection. Examples are oncoming traffic at a crossroads or traffic coming from the side streets at a crossroads.
    \item \textbf{Pedestrian and non-motorized vehicle interactions:} in these scenarios the ADS has to deal with pedestrians and non-motorized vehicles participating in traffic. Examples include pedestrians crossing at a crosswalk, bicycles traveling along the road, and bicycles crossing diagonally at an intersection.
    \item \textbf{Surrounding vehicle interactions:} these scenarios deal with vehicles traveling in the same direction as the ADS. Examples of this are vehicles merging in the lane ahead, vehicles in front slowing down to a halt before accelerating again, and vehicles joining the lane from an access ramp.
    \item \textbf{Emergency evasion:} these are emergency scenarios in which the ADS needs to take urgent action. Examples are emergency braking after a vehicle merges in front, sudden pedestrians crossing, and adjacent vehicles losing control.
\end{itemize}

\begin{figure}
    \centering
    \includegraphics[width=\linewidth]{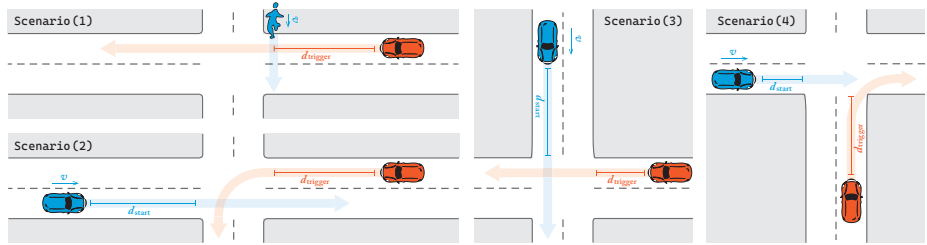}
    \caption{Example scenario: a pedestrian crossing in front of the ADS.}
    \label{fig:scenex}
\end{figure}

Four examples of scenarios in ISS-Scenario can be seen in \Cref{fig:scenex}. Consider \emph{Scenario(1)}, where the ADS encounters a pedestrian crossing in front of the vehicle. Two possible parameters of this scenario are $v$, the velocity of the pedestrian, and $d_{\textrm{trigger}}$, the distance between the car and the pedestrian when the pedestrian starts crossing the road.

\section{Batch Scenario Testing}\label{sec:batch}
\begin{figure}
    \centering
    \includegraphics[width=.666\linewidth]{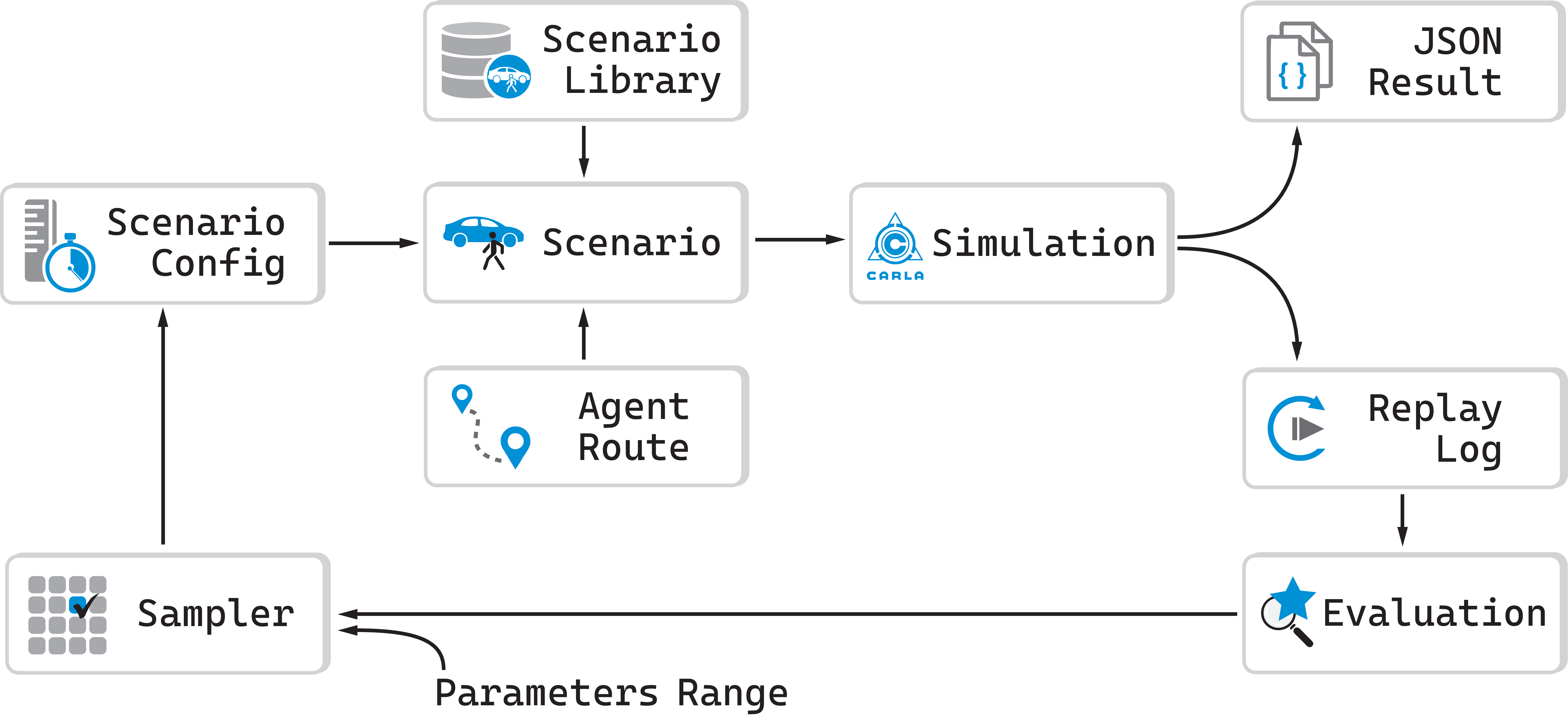}
    \caption{The batch testing pipeline.}
    \label{fig:batchpipe}
\end{figure}

In addition to the scenario library, ISS-Scenario contains a batch testing framework. The basic pipeline for the batch testing procedure is given in \Cref{fig:batchpipe}. The basic procedure has six steps:
\begin{enumerate}
    \item \textbf{Scenario configuration:} batch testing starts by delineating the circumstances in which the ADS will be tested. Different parameters such as "cloudiness" and "pedestrian velocity" are given a range of possible values with which the ADS will be tested.
    \item \textbf{Sampling:} the sampling module is responsible for generating specific scenario configurations by fixing parameter values within their possible ranges. There are two different sampling methods: \emph{uniform sampling,} where the values are picked uniformly at random within the range, and \emph{optimization-based sampling,} which uses a genetic algorithm to search for parameter values that lead to interesting scenarios.
    \item \textbf{Scenario generation:} once the parameter values are sampled, a specific scenario configuration file is generated using the sampled parameter values.
    \item \textbf{Simulation:} the ADS is then tested against this specific scenario in CARLA.
    \item \textbf{Evaluation:} after completion of the scenario, a log file and a JSON file are stored containing the outcomes of the simulation. The evaluation module assesses the performance of the ADS using the log file. Evaluation criteria include the minimum distance between the ADS and other traffic participants, with a higher score being assigned for a smaller distance, indicating a close encounter or potential collision.
    \item \textbf{Iteration:} the process restarts at the sampling step, with the sampler's decisions potentially being guided by the previous evaluation.
\end{enumerate}
The replay log coming out of the simulation step can be used to deterministically rerun specific simulations, allowing for comprehensive analysis of accidents using all of the simulator data.
 
\begin{figure}
\begin{tikzpicture}
\begin{axis}[
width=0.8\linewidth,
height=0.3\linewidth,
ybar,
enlargelimits=0.15,
legend style={
    legend cell align=left,
    legend columns=2,
    at={(0.5,-0.28)},anchor=north,},
ylabel={\# Collision cases},
ylabel style={font=\large},
xtick=data,
xticklabels={i,ii,iii,iv}
]
\addplot[blue,fill=blue!40!white]
coordinates{
   (1,215)
   (2,199)
   (3,213)
   (4,197)
};

\addplot[blue,fill=blue!70!white]
coordinates{
   (1,1960)
   (2,1857)
   (3,1562)
   (4,905)
};
\addplot[red,fill=red!40!white,error bars/.cd,y dir=both,y explicit,]
coordinates {
  (1,173)
  (2,230)
  (3,253)
  (4,683)
};
\addplot[red,fill=red!70!white,error bars/.cd,y dir=both,y explicit,]
coordinates {
  (1,1960)
  (2,1751)
  (3,2509)
  (4,2348)
};
\legend{TCP (random),TCP (genetic),Interfuser (random), Interfuser (genetic)}
\end{axis}
\end{tikzpicture}
\caption{Number of collision cases found in 4 scenarios. Compared to random search, the genetic algorithm was able to find many more collisions.}
\label{table:case_study_results}
\end{figure}
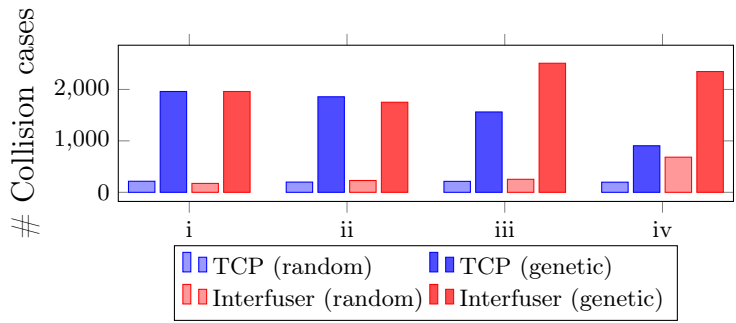

\section{Case studies}\label{sec:case}
We evaluate the testing framework by examining two end-to-end autonomous driving systems, namely TCP and Interfuser. For this case study, we employ four distinct scenarios to investigate potential collision cases using both random sampling and genetic algorithms. The selected scenarios are depicted in Fig.~\ref{fig:scenex}. The scenario parameters are the start distance, the trigger distance and the actor velocity. All other parameters are kept constant. All experiments are conducted on two servers equipped with AMD EPYC 7543 CPUs, 128GB of RAM, and 4 Nvidia RTX 3090 GPUs.

\paragraph{\textbf{Experimental Results:}} We present the experimental results in \Cref{table:case_study_results}, indicating the total number of generated testing cases and the identified collision cases. Regarding testing efficiency, a notable improvement is observed in the genetic algorithm approach, resulting in the discovery of more collision cases. The significant proportion of collision cases underscores the challenge that even end-to-end ADSs with robust deep learning algorithms may struggle to handle complex traffic corner cases effectively.

\begin{figure}
    \centering
    \includegraphics[width=.666\linewidth]{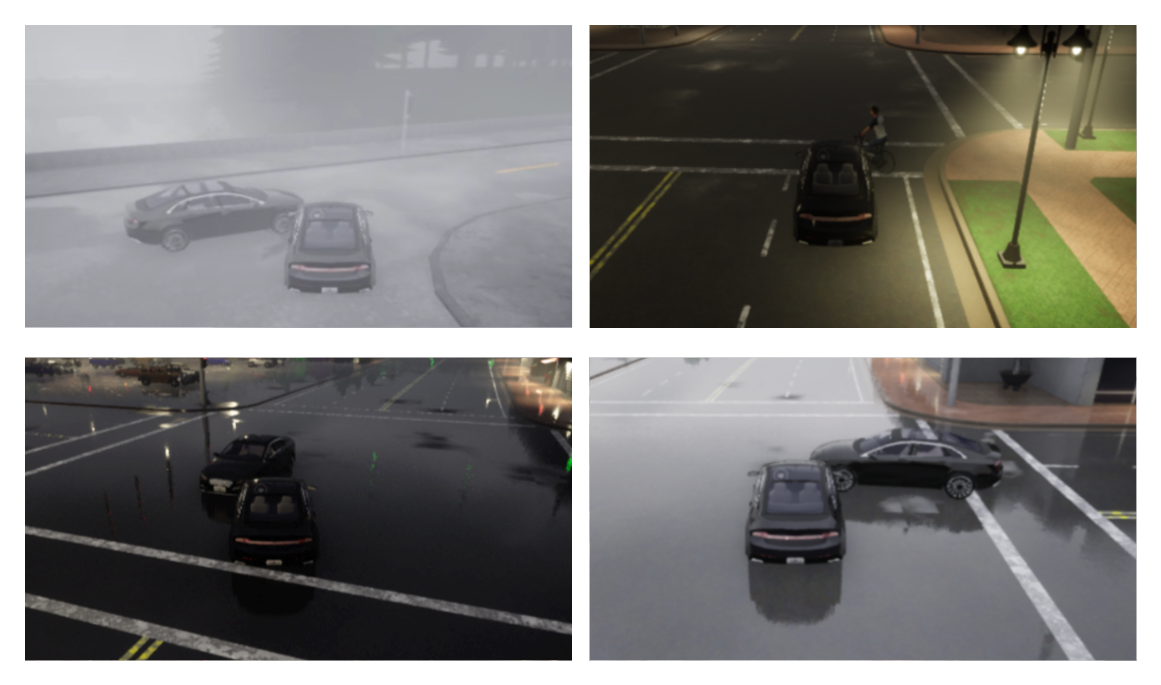}
    \caption{Replay of collision cases. Above: TCP Below: Interfuser}
    \label{fig:case_study_running_examples}
\end{figure}

\paragraph{\textbf{Collision Cases Analysis:}} We further investigate the collision cases. The accident replay feature enables us to deterministically rerun the simulations that lead to these collision cases. We illustrate some examples of these in \Cref{fig:case_study_running_examples}. We found that the ADSs are more prone to be unsafe during the night time.

\section{Conclusion}\label{sec:conc}
In this paper we introduced ISS-Scenario. ISS-Scenario provides an extensive, parametrized scenario library for the evaluation of autonomous driving systems. Moreover, it contains a framework for the automated testing of an ADS against many different settings including a genetic algorithm to search intelligently for potentially dangerous settings. We evaluated ISS-Scenario by testing the ADSs TCP and InterFuser against four different scenarios, showing that these ADSs are more prone to unsafe behavior during the simulated night time. ISS-Scenario is available at \url{https://github.com/CAS-LRJ/ISS\_Scenario}.

\subsubsection{Acknowledgements} This study is part of the CAS Project for young Scientists in Basic Research (YSBR-40) and ISCAS New cultivation Project (ISCAS-PYFX-202201). We are grateful to the National Natural Science Foundation of China (62172217) and the Chinese Academy of Sciences for funding this work.

\protect\includegraphics[height=8pt]{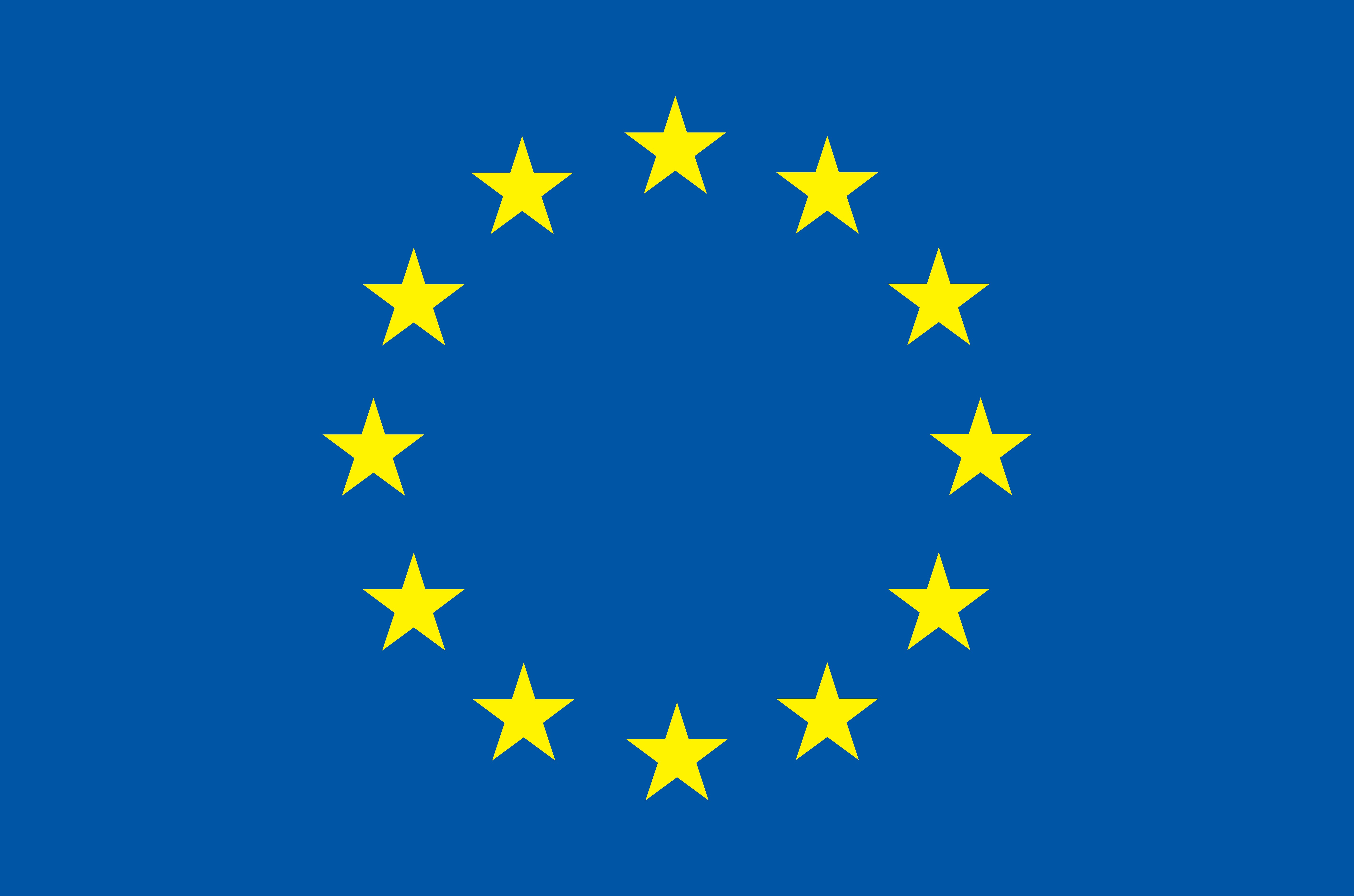} This work is part of the European Union’s Horizon 2020 research and innovation programme under the Marie Sk\l{}odowska-Curie grant no.\@ 101008233.

%
%
%
\bibliographystyle{splncs04}
\bibliography{main}
%




\end{document}